\documentclass[aps,prl,twocolumn,a4paper,10pt,notitlepage,footinbib,superscriptaddress,showpacs]{revtex4-1}%
\usepackage[english]{babel}
\usepackage[T1]{fontenc}
\usepackage{endnotes}
\usepackage{amssymb,amsmath,amsfonts}
\usepackage{textcomp}
\usepackage{graphicx,color}
\usepackage[utf8x]{inputenc}
\usepackage{float}
\usepackage{placeins}
\usepackage{multirow}

\usepackage{subfigure}

\begin{document}

\title{Multi-scale control of active emulsion dynamics}

\date{\today}

\author{Livio Nicola Carenza}
\affiliation{Dipartimento  di  Fisica,  Universit\`a  degli  Studi  di  Bari  and  INFN,  via  Amendola  173,  Bari,  I-70126,  Italy}

\author{Luca Biferale}
\affiliation{Dipartimento di Fisica and INFN, Universit\`a di Roma “Tor Vergata”, Via Ricerca Scientifica 1, 00133 Roma, Italy}

\author{Giuseppe Gonnella}
\affiliation{Dipartimento  di  Fisica,  Universit\`a  degli  Studi  di  Bari  and  INFN,
Sezione  di  Bari,  via  Amendola  173,  Bari,  I-70126,  Italy}

\begin{abstract}
We numerically study the energy transfer in a multi-component $2d$ film, made of an active polar gel and a passive isotropic fluid in presence of surfactant favoring emulsification. We show that by confining the active behavior into the localized  component, the typical scale where chemical energy is transformed in mechanical energy can be substantially controlled. Quantitative analysis of kinetic energy spectra and fluxes shows the presence of a multi-scale dynamics  due to the existence of a flux induced by the active stress only, without the presence of a turbulent cascade. An increase in the intensity of active doping induces drag reduction due to the competition of elastic and dissipative stresses against active forces. Furthermore we show that a non-homogeneous activity pattern  induces  localized response, including a modulation of the slip length, opening the way toward the control of active flows by external doping.
\end{abstract}

\maketitle


Active fluids exhibit a number of peculiar behaviors due to the small-scale conversion of internal into mechanical energy by the active constituents.
Many biological examples, such as bacterial~\cite{wensink2012,dunkel2013} and cytoskeletal suspensions~\cite{sanchez2012,guillamat2016,guillamat2018} and artificial systems, e.g. Janus~\cite{ebbens2014,gregory2015} and magnetic microparticles~\cite{kokot2017},  have been studied in the labs and by numerical simulations. In the presence of high concentration of the active component, there exists the possibility to develop  complex (chaotic) flows even in absence of any external forcing, due to active injection only~\cite{marchetti2013,Doostmohammadi2017,carenza2019}.
This is a non-trivial collective phenomenon \cite{voituriez2005}, also leading to rich rheological properties, including cases of vanishing and negative effective viscosity~\cite{cates2008,lopez2015,loisy2018,negro2019}, or preferential  clustering~\cite{fily2012,lowen2013,digregorio2018}.\\
\begin{figure*}[!bt]
\centering
\includegraphics[width=1.\textwidth]{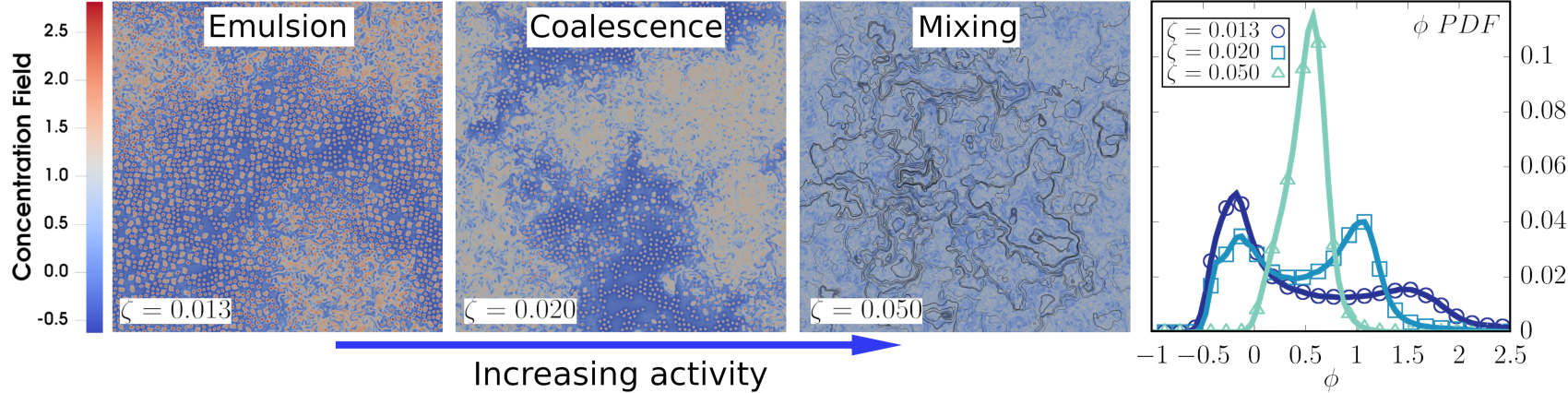} 
\caption{Contour-plot of concentration field $\phi$, where the active component is  red and passive is blue showing 
late-time configurations of active fluids at different intensities of active doping $\zeta$. Velocity streamlines are plotted in black for the case at $\zeta=0.050$. Last panel:  $pdf$ of the concentration field,
notice the transition to mixing for $\zeta>0.020$ characterized by the presence of a single peak.
}
\label{fig:fig1}
\end{figure*} 
Understanding the dense-suspension limit is pivotal to develop novel fluid materials with \emph{ad-hoc} and controllable space-time dependent features, a challenge for both fundamental and applied (micro) fluid mechanics.  Recently, very interesting results have been produced, pointing toward the possibility to develop fluid motion at meso-scales, \emph{i.e.}  at distances much larger than the typical single-agent size and with a kinetic energy spectrum characterized by power-law behaviour in a limited range of scales. The term {\it bacterial turbulence} has been coined for that  \cite{Bratanov2015,dombrowski2004,Doostmohammadi2017,Shendruk2017,giomi2015,wensink2012}, with a puzzling variety of non-universal behaviours \cite{wensink2012,Bratanov2015,giomi2015,creppy2015,kokot2017,Linkmann2019}.\\
In this paper, we want to quantitatively disentangle the different mechanisms behind the development of complex active flows, for the important case of a composite fluid: a $2d$  active polar gel in a passive isotropic fluid matrix \cite{bonelli2019,negro2018}. At difference from most of the previous studies,  we are interested in the set-up where the active matter is confined in a droplet-like emulsion phase (see Fig. 1), whose experimental realization can be achieved, e.g.,   by confining cellular extracts in a water-in-oil emulsion~\cite{guillamat2016,sanchez2012} or in bacterial systems subjected to depletion forces leading to microphase separation~\cite{schw2012}.  Our set-up  is particularly appealing because it allows us  to change the degree of localization of the energy injection by the active matter -- a novel way to control forcing mechanisms in real flows. 
By moving from a confined emulsion to a phase with large aggregates of active material and then to a fully dispersed (mixed) regime, we are able to systematically address the relevant scales of the active-component 
for the chaotic flow evolution (see Fig. 1). We clarify the difference between  an {\it active turbulent} and an  {\it active elastic} flow and we argue that in most cases, one cannot speak of a turbulent non-linear cascade,  being the velocity field driven by the active field only. We discuss what are the key  observable that must be controlled in order to distinguish and classify the flow properties, stressing that the energy spectrum is not informative enough and cannot be used to distinguish between the presence of an inverse/direct cascade or even no-cascade at all~\cite{alexakis2018}. Concerning more applied aspects, we show that the flow has a non-trivial global response at changing the doping of the confined active phase, with a tendency to reduce the drag by going toward more and more delocalized agents. 
Finally, we show that a suitable space-time modulation of the doping  is capable to fine-tune the flow response and the mixture morphology, opening the unexplored direction  for active-control of (micro) active fluids.\\
{\sc The Model.} The orientable nature of many active constituents has been successfully modeled by means of the Landau-De Gennes theory for liquid-crystals, introducing a coarse-grained \emph{polarization field} $\textbf{P}$, accounting for the local orientation of constituents, while the local \emph{concentration} of active material is described by the conserved field $\phi$.
The evolution of the system is governed by the following equations, in the limit of incompressible flow:
\begin{equation}
    \begin{cases}
\rho \left(\partial_t + \textbf{v}\cdot \nabla\right) \textbf{v} =\nabla \cdot \left( \tilde{\sigma}^{pass} +\tilde{\sigma}^{act} \right), \\
\partial_t \phi+\nabla\cdot\left(\phi\mathbf{v}\right)=M \nabla^2 \mu, \\
\partial_t \mathbf{P}+\left(\mathbf{v}\cdot\nabla\right)\mathbf{P}=-\tilde{\Omega}\cdot\mathbf{P}+\xi\tilde{D}\cdot\mathbf{P}-\frac{\textbf{h}}{\Gamma} .
\label{eqn:dynamical_eq}
\end{cases}
\end{equation}
Here the first equation is the incompressible Navier-Stokes equation, where $\textbf{v}$ is the velocity field and $\rho$ the total (constant) density, while the stress tensor has been divided in a passive term, $$\tilde{\sigma}^{pass} = \tilde{\sigma}^{hydro} + \tilde{\sigma}^{bin} + \tilde{\sigma}^{pol},$$
given by the conserved momentum current, including viscous and ideal fluid pressure contributions (see SM), and a phenomenological traceless active term~\cite{pedley1992,hatwalne2004} $$\tilde{\sigma}^{act}= -\zeta \phi \left(\textbf{P} \otimes \textbf{P} - \dfrac{\textbf{I}}{2} \textbf{P}^2 \right).$$
The \emph{activity} parameter $\zeta$ tunes the intensity of active doping: if positive, it describes the stress exerted by pusher swimmers -- thus generating \emph{extensile} quadrupolar flow patterns~\cite{hatwalne2004}. Pullers can be modeled with negative values of $\zeta$. In this Letter, we will restrict to $\zeta>0$, since most of biological extracts exhibiting complex behaviors are pushers.
The second equation rules the convection-diffusion  evolution of active concentration $\phi$. Here $M$ is the mobility, while the  chemical potential $\mu={\delta \mathcal{F}}/{\delta \phi}$ is derived from a generalization of the Brazovskii free energy~\cite{braz1975,gonnella1997,bonelli2019,negro2018}:
\begin{multline}
    \mathcal{F} \left[ \phi, \mathbf{P} \right] = \int  \left[ \dfrac{a}{4 \phi_{cr}^2} \phi^2 (\phi - \phi_0)^2 + \dfrac{k_\phi}{2} (\mathbf{\nabla} \phi)^2 + \dfrac{c}{4} (\mathbf{\nabla}^2 \phi)^2 \right. \\ \left. -\dfrac{\alpha}{2} \dfrac{(\phi-\phi_{cr})}{\phi_{cr}} \mathbf{P}^2 + \dfrac{\alpha}{4} \mathbf{P}^4 + \dfrac{k_P}{2} (\mathbf{\nabla} \mathbf{P})^2 + \beta \mathbf{P} \cdot \mathbf{\nabla} \phi \right] \text{d} \mathbf{r}.
\label{eqn:freeE}
\end{multline}
Phase separation of the two components is obtained with bulk energy density $a>0$, to have two minima in the free-energy at $\phi=0,\phi_0$. By choosing  $k_\phi<0$, interface formation is favored, so that the Brazovskii constant $c$ must be positive to guarantee thermodynamic stability. 
Setting $\alpha>0$, the polarization field $\mathbf{P}$ is confined in the \emph{active} regions (where $\phi>\phi_{cr}=\phi_0/2$), and is absent in \emph{passive} regions, where $\phi<\phi_{cr}$. The energy cost for the elastic deformations is paid by the gradient term $(\mathbf{\nabla} \mathbf{P})^2)$~\cite{degennes1993}. The coupling $\mathbf{P}\cdot \nabla \phi$ defines the anchoring of the vector field at interfaces. If $\beta>0$ the polarization at interfaces will point towards passive regions of the mixture.
In the passive limit ($\zeta=0$) and for asymmetric compositions ($ \overline{\phi/\phi_0} \lesssim 0.35$, with the bar denoting spacial average) the system sets into an array of droplets of the minority phase in a background of the majority phase~\cite{negro2019}, see also Fig.~\ref{fig:fig1}. 
Finally, the evolution of the polarization field is governed by the Ericksen-Leslie equation for a vector order parameter. Here $\textbf{h}={\delta \mathcal{F}}/{\delta \textbf{P}}$ is the molecular field and $\Gamma$ the rotational viscosity, while the symmetric deformation rate tensor is $\tilde{D}=\frac{1}{2}(\nabla \textbf{v} + \nabla \textbf{v}^T)$ and the vorticity tensor is $\tilde{\Omega}=\frac{1}{2}(\nabla \textbf{v} - \nabla \textbf{v}^T)$; we choose the shape factor $\xi>1$ to model flow-aligning rod-like swimmers~\cite{bonelli2019,negro2018}.
In Fig. 1 we show for the first time 
the existence of a transition to a final mixed phase by increasing the activity $\zeta$. This is due to the fact that strengthening the doping, induces more and more active stress  across the droplets leading to interface breaking and droplet coalescence. 
The consecutive dispersion of active agents in the whole volume has the important effect to change the typical flow length-scales since small-scale deformation of the polarization pattern are wiped out. Similarly, a decrease in the viscosity $\eta$ makes more efficient the active pumping in the flow, leading to the same conclusion. In what follows, we will  investigate the kinematic and flow properties across this transition. \\
{\sc Numerical Results.} We performed a systematic series of numerical simulations by fixing all free parameters in Eqs.~\eqref{eqn:dynamical_eq} and~\eqref{eqn:freeE} to values that correspond to a realistic description 
of cytoskeletal filaments~\cite{tjhung2012,tjhung2015,sanchez2012} (see Table I and SM) changing the intensity of the active doping $\zeta$, integrating Eqs.~\eqref{eqn:dynamical_eq} by means of a widely validated hybrid lattice Boltzmann (LB) approach, on a squared periodical $2d$-lattice of size $L=1024$, except otherwise stated.
Details about the numerical scheme can be found in the SM. \\
\begin{table}[b]
\caption{Mapping between physical and simulation units. Length-scale $l^*= 1  \mathrm{\mu m}$, time-scale $t^*=10 \mathrm{ms}$ and force-scale $f^*=10^2 \mathrm{nN}$ are fixed to be $1$ LB units. Viscosity $\eta$ is expressed in $\mathrm{kPa}\mathrm{s}$, the elastic constant of the polar gel $k_P$ in $\mathrm{nN}$, the diffusivity $D=Ma$ in $\mu \mathrm{m}^{2}\mathrm{s}^{-1}$, while $\Gamma$ and $\zeta$ are respectively expressed in $\mathrm{kPa}\mathrm{s}$ and $\mathrm{kPa}$.}
\label{table1}
\vskip 0.3cm
\begin{tabular}{l|l|l|l|l|l}
Units  & $\eta$  & $k_P$ & $D$  & $\Gamma$  & $\zeta$ \\
  \hline
 Simulation  & $5/6 \quad$   & $0.01 \quad$ \ \ & $0.0004 \quad$ \ \  & $1 \quad$ \ \  &  $0.01-0.06$ \\
 Physical    & $0.83$  & $10$  & $0.004$  & $10$ &  $100-600$
\end{tabular}
\end{table}
{In Fig.~(\ref{fig:fig2}) we start by showing the energy spectra per unit density $E(k) =0.5 \langle |\textbf{v}(\textbf{k})|^2 \rangle$, 
where $\langle \cdot \rangle $ stands for the steady state spherical average,
for different values of $\zeta$ (left panel). 
Remarkably, we found the presence of a continuum spectrum even for wavenumbers smaller than the typical injection scale $k_a$, where the active matter is preferentially confined. The spectrum shape depends on the activity $\zeta$, with an accumulation of energy at large length-scales  for increasing activity. 
The typical  Reynolds numbers are $Re \sim 10^{-2}$ to $1$, a regime close to experimental observations, where we defined $Re =\rho E^{1/2}(k_v) l_v/ \eta$ in terms of the typical flow wave-number  $ k_v =  \left(\int_{0}^{\infty} \text{d}k' k' E(k') \right)/ E_{tot}, $ and length-scale $l_v= L/k_v$, where $E_{tot}=\int_{0}^{\infty} \text{d}k' E(k')$ is the total kinetic energy. 
Since deformations both in the polarization and concentration patterns are acting as a source of mechanical energy, the spectrum behavior is to be related with the morphology of the system.  The right panel of the same figure shows the spectral properties of the active energy injection:
\begin{equation}
\mathcal{S}^{act}(k,t) = \langle \textbf{v}^*(\textbf{k},t) \cdot \textbf{F}^{act}(\textbf{k},t) \rangle
\label{eqn:power_spectrum_definition}
\end{equation}
where $\textbf{F}^{act}(k,t)= 2\pi i \textbf{k} \cdot \tilde{\sigma}^{act}(\textbf{k},t)/L$.
As anticipated, energy pumping is considerably localized at high wavenumber ($k_a/L  = l_a^{-1} \simeq 0.1 $)  when activity is low enough, due to small-scale deformations of the polarization field homeotropically and strongly anchored to interfaces~\cite{bonelli2019,negro2018}. By increasing $\zeta$, as the system undergoes first coalescence ($\zeta=0.02$), then mixing, interfaces broaden and progressively disappear, thus attenuating energy supply at high wavenumbers and leading to a situation where energy is injected on a wider range of scales, typical of the active-driven bending instabilities of the polar gel~\cite{giomi2008,voituriez2005}.
Going back to the spectra, we can see that  as long as the droplet phase survives, energy is accumulated on the typical length-scale of the droplet size ($l_d \simeq 15$), as suggested by the bulge in energy spectra, for small values of $\zeta$, located at wavenumber $k_d/L = l_d^{-1} \simeq 0.067$. 
As confinement is lost ($\zeta \gtrsim 0.030$), energy is instead spread on much larger scales ($k_v/L = l_v^{-1} \simeq 0.025 $). 
Even more interesting,  is the observation that the energy injected in the system at low activity is substantially greater than the one delivered at higher active dopings.\\
}
\begin{figure}[t]
\centering
\includegraphics[width=1.\columnwidth]{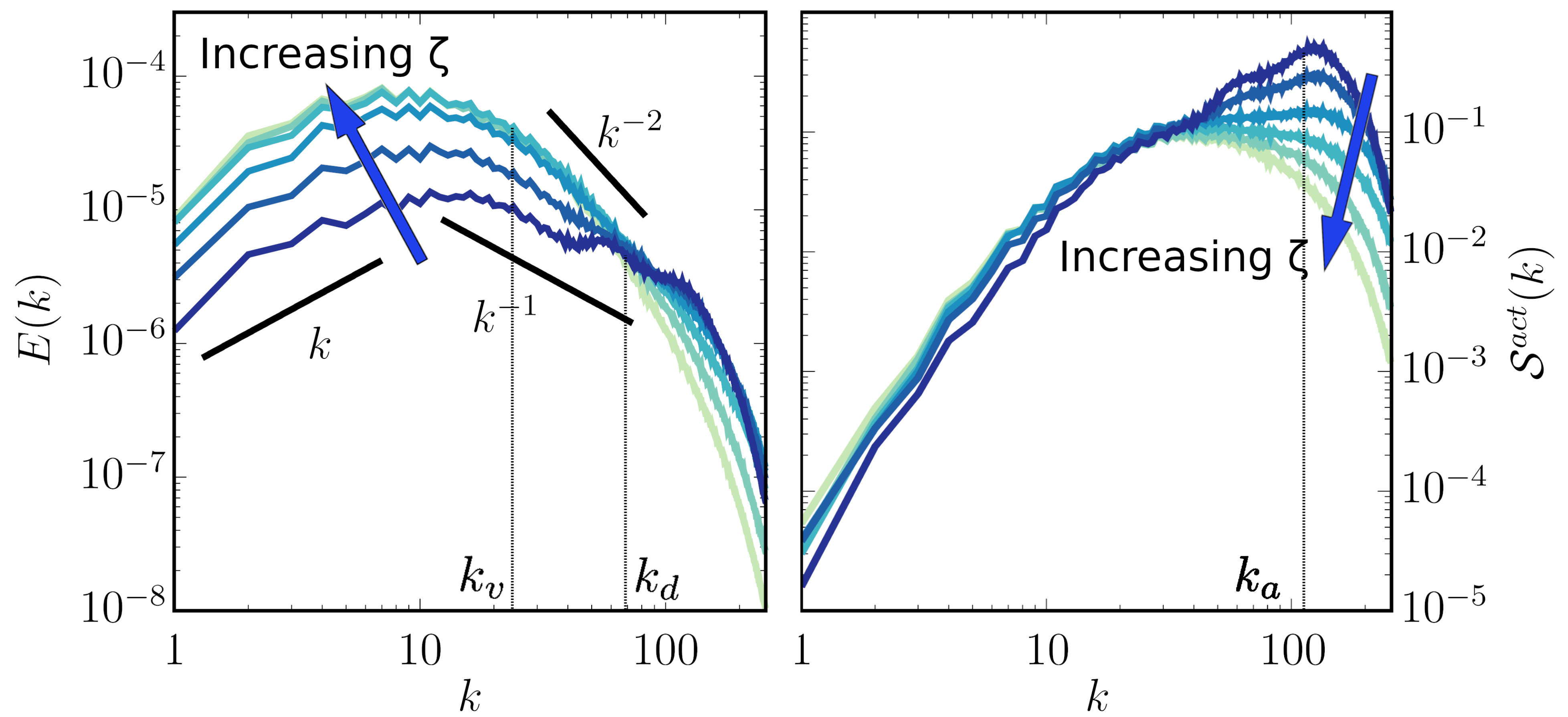}
\caption{{Left: log-log plot of time-averaged energy spectra varying activity ($\zeta=0.013,0.015,0.02,0.03,0.04,0.05$). Right:  total amount of energy injected in the system by active agents. Vertical black dashed  lines mark the wavenumber $k_v,k_d$ and $k_a$ respectively (see text).} }
\label{fig:fig2}
\end{figure}
{This surprising behavior (and its connection to morphology) has been confirmed by keeping fixed the active parameter $\zeta=0.015$ and varying the nominal viscosity of the suspension (not shown). Once again we found that total kinetic energy rises and develops on progressively bigger length-scales as viscous effects are lowered and mixing of the two components occurs, thus driving the active agents in an unconfined state. 
}\\
It is well known that spectra do not bring enough information  to disentangle the complicated network of transfer mechanisms inside a complex flow~\cite{alexakis2018}. 
Thus, we performed a systematic analysis of the energy balance in Fourier space, by looking at the scale-by-scale contribution of all terms, either dissipative or reactive. We thus Fourier-transform both sides of the Navier-Stokes equation and we multiply them by $\textbf{v}^*(\textbf{k})$, to obtain the following balance equation,  spherically averaged on shells of equal momentum:
\begin{equation}
\label{eqn:energy_balance_kspace}
\rho \partial_t E(k,t) + \mathcal{T}(k,t) = \sum_i \mathcal{S}^{(i)}(k,t).
\end{equation}
Here $\mathcal{T}(k,t)= \langle \mathbf{v}^*(\mathbf{k},t) \cdot \mathbf{J}(\mathbf{k},t) \rangle$ represents the rate of energy transfer due to nonlinear hydrodynamic interactions (with $\mathbf{J}(\mathbf{k},t)$ standing for the Fourier transform of $\rho \mathbf{v} \cdot \mathbf{\nabla} \mathbf{v}+ \mathbf{\nabla} p$). The terms on the right-hand side of Eq.~\eqref{eqn:energy_balance_kspace} are energy source/sink contributions, where the terms $\mathcal{S}^{(i)}(k,t)$ are defined as in Eq.~\eqref{eqn:power_spectrum_definition} and $(i)$ denotes the different kinds of contributions (viscous, binary, polar or active).
Finally we define each separate component of the energy flux as the total variation per unit time of the energy contained in a sphere of radius $k$: $$ \Pi^{(i)}(k,t) = \int_0^k \text{d}k'~ \mathcal{S}^{(i)}(k',t),$$ with $\Pi_\mathcal{T}(k,t)$ defined analogously.
Fig.~\ref{fig:fig3} shows 
fluxes for $\zeta=0.013,0.050$, measured at 
steady state. In both cases, the only source contribution is the active one, $\Pi^{act}$ while all the others are energy sinks.
Before the transition to mixing ($\zeta=0.013$ left panel) the binary and polar terms are also contributing (as sinks). After, the dynamics is characterized by an almost perfect matching among viscous and active terms (right panel). The advection term, $\Pi_\mathcal{T}$,  is, for all practical effects, null, as expected for fluids flowing at negligible Reynolds numbers. 
The previous findings suggest some important conclusions. First, the scenario is in agreement with {\it the absence} of hydrodynamic turbulence.
Here, multi-scale effects and chaotic evolution  arise  from the competition between sink terms and active injection leading to a non trivial scale-to-scale balance.
The phenomenology is similar to the case of {\it elastic turbulence}, where the non-linear  evolution for the velocity field is dominated  by the non-Newtonian stress, leading to chaotic and unpredictable multi-scale evolution even at nominally vanishing Reynolds numbers~\cite{Larson2000,Burghelea2006,morozov2007}. Second, the overwhelming role played by the active stress explains the absence of universality~\cite{wensink2012,Bratanov2015,kokot2017} in many bio-fluids: there is no direct or inverse energy cascade mediated by the universal advection flux, $\Pi_\mathcal{T}$. Energy is moved from scale to scale by the direct interaction with the case-specific active  stress. The  absence of a turbulent cascade is also supported by the relatively small extension of the wave-range where fluxes are non-zero, as seen from Fig.~\ref{fig:fig3}. Indeed, the vanishing of all $\Pi^{(i)}$ for $k <10$ leads to a quasi-equilibrium range at large scale, where the spectrum $E(k) \sim k$ (see Fig.~\ref{fig:fig2}). \\
\begin{figure}[t]
\centering
\includegraphics[width=1.\columnwidth]{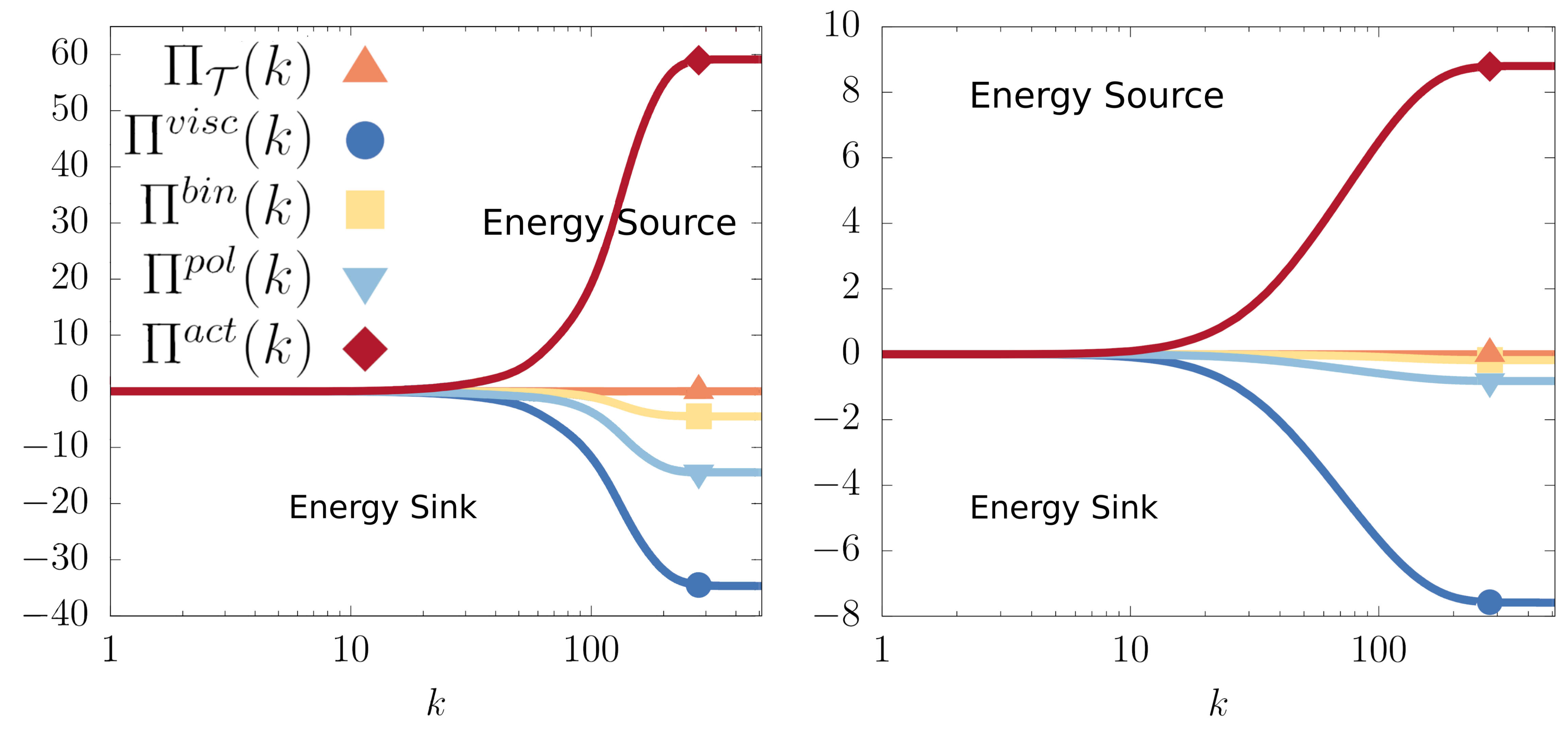}
\caption{Time-averaged components of the  total energy flux at $\zeta=0.013$ (left panel) and $\zeta=0.050$ (right panel). 
Notice the different y-range in the two graphs.
}
\label{fig:fig3}
\end{figure}
\begin{figure}[b]
\centering
\includegraphics[width=1.\columnwidth]{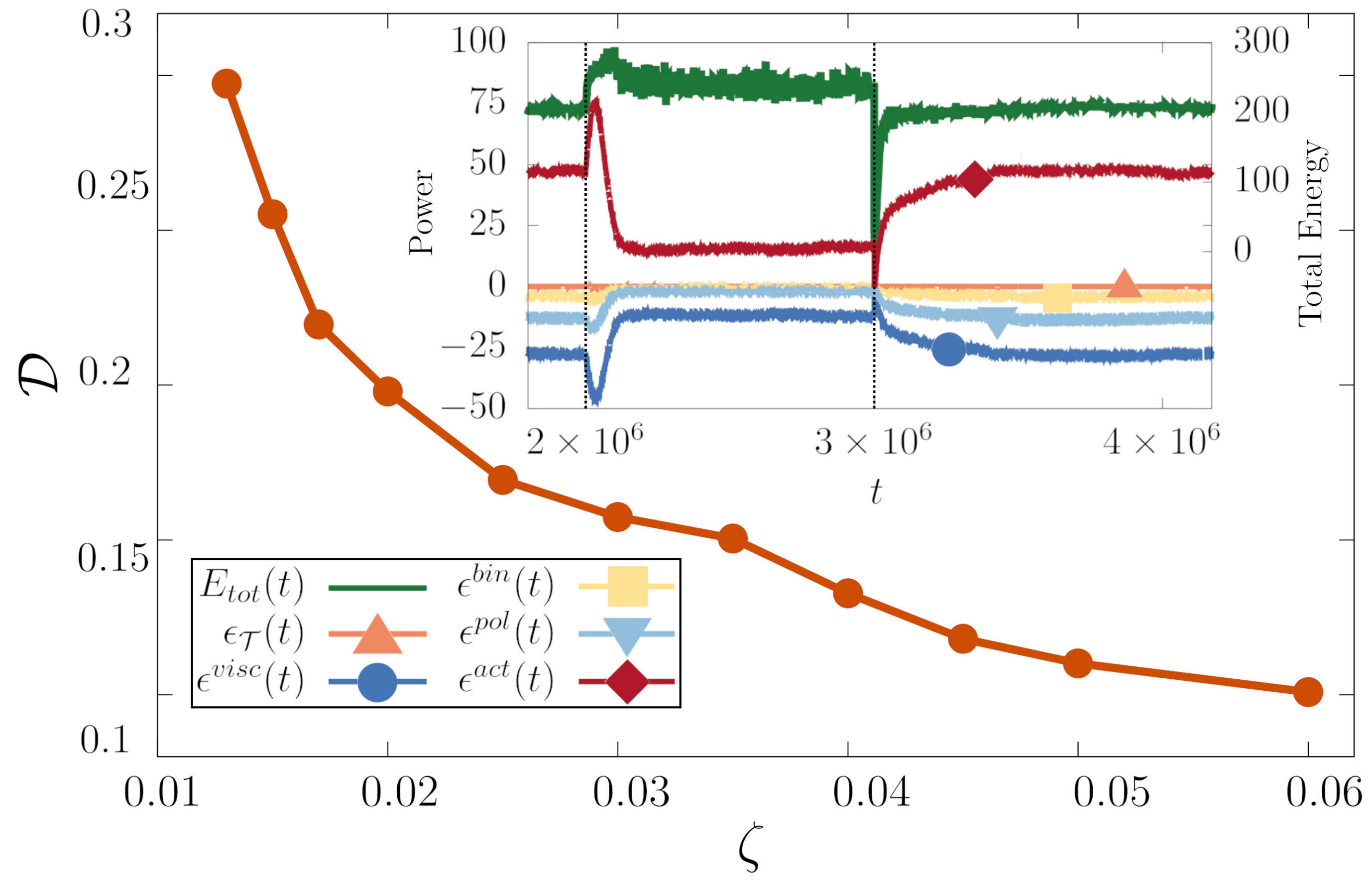}
\caption{Drag factor versus activity. 
Inset shows time evolution of kinetic energy $E_{tot}$ and power terms $\epsilon^{(i)}(t)$ defined in the same way as $\epsilon^{act}(t)$, while rising activity from $\zeta=0.010$ to $\zeta=0.030$ at time $t_1$ and reducing it to its previous value at time $t_2$, denoted by vertical dashed lines. }
\label{fig:fig4}
\end{figure}
To quantitatively characterize the efficiency of injection of energy pumped in the system by active effects,
we define the \emph{drag factor}: $$\mathcal{D}=\frac{\epsilon^{act} l_{v}}{E^{3/2}_{tot}} $$ in terms of the total energy injected by active forces, the only source contribution
$\epsilon^{act} = lim_{k\to \infty} \Pi^{act}(k)$,  the total response, given by the available  kinetic energy $E_{tot}$, and  $l_v$. 
In Fig.~\ref{fig:fig4} we show the results for $\mathcal{D}$ at varying the activity $\zeta$. 
We note that the drag factor rapidly decreases while increasing activity in the emulsion phase, then the decrease slows down as big active clusters become dominant in the system, and finally tends to saturation when the morphological transition towards mixing takes place.\\
{\sc Space-time control}. Finally, to test the dynamical response of active emulsions we performed a numerical study when we abruptly rise $\zeta$ across the mixing transition, from $0.013$ to $\zeta=0.030$. Kinetic energy rapidly increases in a short transient, and finally sets to a new stable value higher than before. Power terms behavior is instead characterized by a spike in correspondence of the transient, followed by relaxation towards much smaller values. This behavior is found to be reversible: indeed when activity is lowered again kinetic energy rapidly restore its previous values, as well as power terms (see inset of Fig.~\ref{fig:fig4}). Finally, in Fig.~\ref{fig:fig5} we show a space-dependent protocol to control the flow response. We studied the evolution of the active polar gel emulsion on a channel under constant pressure gradient and by changing the activity parameter as a function of the position downstream $\zeta(x)$. 
\begin{figure}
\centering
\includegraphics[width=1.\columnwidth]{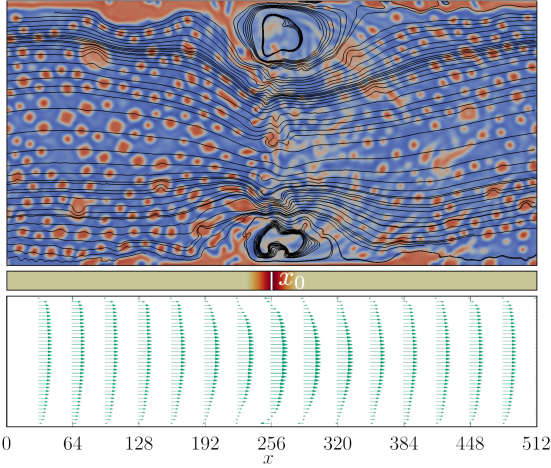}
\caption{Contour plot of concentration of active and passive phases in an externally controlled active emulsion subject to Poiseuille flow in a $512 \times 256$ channel (color legend is the same as in Fig.~\ref{fig:fig1}). Velocity streamlines are plotted in black, showing effective reduction of the channel width. Activity $\zeta$ is modulated in the flow direction as shown in the central color bar (beige corresponds to $\zeta=0.010$, while $\zeta(x_0)= 0.03$. Time-averaged profiles of $\textit{v}_x$ are shown in the last panel. Notice that profiles close to $x_0$ exhibit a non-null (negative) slip length.}
\label{fig:fig5}
\end{figure}
In particular we show that a local sharp increase of $\zeta$ above the mixing transition around $x \sim x_0$ is an efficient tool to locally control the degree of emulsification in the bulk, opening the way to change the flow transport properties and the flow topology {\it on-the-fly}.  \\
{\sc Conclusions.} 
By using systematic numerical investigation of a $2d$ active emulsion we have shown that a non-trivial multi-scale chaotic dynamics develops due to the 
direct injection of chemical/internal energy in to the flow evolution. We showed that the relative effects of advection, viscous, reactive forces do change at varying the activity of the dispersed phase. In particular, for low activity, the flow structures are driven by the active stress and dissipated by the polar and viscous components mainly. For large activity, the flow undergoes a morphological  transition, with the two fluids (active and passive) both well mixed and active forces are balanced by the viscous drag only. This must be considered a first hint that active emulsions are able to develop  flow configurations with a wide range of dynamical scales.   An increase in the intensity of active doping induces drag reduction due to the competition of elastic and dissipative stresses against active forces. Furthermore we have shown that a non-homogeneous pattern of activity in the bulk induces a localized response opening the way toward the control of active flows by patterned external doping.

\bibliographystyle{unsrt}
\bibliography{biblio}

\appendix

\clearpage

\section{1. Stress tensor}
The stress tensor appearing at the right-hand side of the Navier-Stokes equation has been divided in two parts, respectively addressed as \emph{passive} $\tilde{\mathbf{\sigma}}^{pass}$ and \emph{active} $\tilde{\mathbf{\sigma}}^{act}$. The first is in turn the sum of dissipative and reactive contributions. The hydrodynamic term, including ideal fluid pressure and viscous dissipation~\cite{landau1987} is given by: $$ \tilde{\mathbf{\sigma}}^{hydro}_{\alpha \beta} = - p \delta_{\alpha \beta} + \eta (\partial_\beta v_\alpha + \partial_\alpha v_\beta).$$ 
Poiseuille flow, studied in Fig.~5 of the main text, is obtained by applying a body force $f_x$,
\emph{i.e.} requiring the pressure gradient to satisfy  $\nabla p=-f_x$. 
Reactive terms include equilibrium contributions~\cite{degennes1993} arising from the free energy functional $\mathcal{F} \left[ \phi, \mathbf{P} \right] $ (Eq.~(2) of the main text), which can be in turn divided in a binary mixture term $$\tilde{\sigma}_{\alpha\beta}^{bin}=\left( f-\phi\frac{\delta F}{\delta\phi} \right)\delta_{\alpha\beta} - \frac{\delta \mathcal{F}}{\delta (\partial_\beta \phi)} \partial_\alpha \phi,$$
and in a polarization term $$\tilde{\sigma}_{\alpha\beta}^{pol}= \frac{1}{2}(P_{\alpha}h_{\beta}-P_{\beta}h_{\alpha})-\frac{\xi}{2}(P_{\alpha}h_{\beta}+P_{\beta}h_{\alpha}) -\frac{\delta \mathcal{F}}{\delta (\partial_\beta P_\gamma)} \partial_\alpha P_\gamma,$$
where $\mathbf{h}=\delta \mathcal{F} / \delta \mathbf{P}$ stands for the molecular field~\cite{degennes1993}. $\xi$ is the shape factor and selects rod-like particles if positive or disk-like ones if negative. Moreover, the polarization field exhibits flow aligning or flow thumbling features under shear if $|\xi|>1$ or $|\xi|<1$, respectively.

Energy injection due to the action of active agents is introduced in the model by means of a phenomenological term addressed as \emph{active}: $$ \tilde{\sigma}_{\alpha\beta}^{act}= - \zeta \phi \left(P_\alpha P_\beta -\dfrac{1}{2} |\mathbf{P}|^2 \delta_{\alpha \beta} \right),$$
arising from a coarse-grained description of the force density exerted by pusher ($\zeta>0$) or puller ($\zeta<0$) swimmers on the surrounding fluid~\cite{hatwalne2004,pedley1992}.

\section{2. Numerical methods and parameters}

To solve the hydrodynamics of the system we made use of a hybrid lattice Boltzmann (LB) approach on a $d2Q9$ lattice~\cite{succi2001}. Navier-Stokes equation was solved through a predictor-corrector LB scheme~\cite{denniston2001,bonelli2019,carenza2019}, while the evolution equations for the order parameters $\phi$ and $\mathbf{P}$ were integrated through a predictor-corrector finite-difference algorithm implementing first-order upwind scheme and fourth order accurate stencils for space derivatives.
In this approach the evolution of the fluid is described in terms of a set of distribution functions ${f_i(\mathbf{r}_\alpha,t)}$ (with index $i$ labelling different lattice directions, thus ranging from $1$ to $9$) defined on each lattice site $\textbf{r}_\alpha$. Their evolution follows a discretized predictor-corrector version of the Boltzmann equation in the BGK approximation:
\begin{equation}
f_i (\mathbf{r}_\alpha + \bm{\xi}_i \Delta t) - f_i (\mathbf{r}_\alpha,t) = - \dfrac{\Delta t }{2} \left[ \mathcal{C}(f_i,\mathbf{r}_\alpha, t) + \mathcal{C}(f_i^*,\mathbf{r}_\alpha+ \mathbf{\xi}_i \Delta t, t) \right].
\label{eqn:LBevolution}
\end{equation} 
Here $\lbrace \bm{\xi}_i \rbrace$ is the set of discrete velocities, with $\bm{\xi}_0=(0,0)$, $\bm{\xi}_{1,2}=(\pm u,0)$, $\bm{\xi}_{3,4}=(0,\pm u)$, $\bm{\xi}_{5,6}=(\pm u, \pm u)$, $\bm{\xi}_{7,8}=(\pm u, \mp u)$, where $u$ is the lattice speed. The distribution functions $f^*$ are first order estimations to  $f_i (\mathbf{r}_\alpha + \bm{\xi}_i \Delta t) $ obtained by setting $f_i^* \equiv f_i$ in Eq.~\eqref{eqn:LBevolution}, and $\mathcal{C}(f,\mathbf{r}_\alpha, t)=-(f_i-f_i^{eq})/\tau + F_i$ is the collisional operator in the BGK approximation expressed in terms of the equilibrium distribution functions $f_i^{eq}$ and supplemented with an extra forcing term for the treatment of the anti-symmetric part of the stress tensor.
The density and momentum of the fluid are defined in terms of the distribution functions as follows:
\begin{equation}
\sum_i f_i = \rho \qquad \sum_i f_i \bm{\xi}_i = \rho \mathbf{v}.
\label{eqn:variables_hydro}
\end{equation}
The same relations hold for the equilibrium distribution functions, thus ensuring mass and momentum conservation. 
In order to correctly reproduce the Navier-Stokes equation we impose the following condition on the second moment of the equilibrium distribution functions:
\begin{equation}
\sum_i f_i \bm{\xi}_i \otimes \bm{\xi}_i = \rho \mathbf{v} \otimes \mathbf{v} -\tilde{\sigma}^{bin} - \tilde{\sigma}^{pol}_s,
\label{eqn:constrain_second_moment}
\end{equation}
and on the force term:
\begin{equation}
\sum_i F_i = 0, \qquad \sum_i F_i \bm{\xi}_i = \mathbf{\nabla} \cdot (\tilde{\sigma}^{pol}_a + \tilde{\sigma}^{act}), \qquad \sum_i F_i \bm{\xi}_i \otimes \bm{\xi}_i = 0,
\label{eqn:constraint_force}
\end{equation}
where we respectively denoted with $\tilde{\sigma}^{pol}_s$ and $\tilde{\sigma}^{pol}_a$ the symmetric and anti-symmetric part of the polar stress tensor.
The equilibrium distribution functions are expanded up to the second order in the velocities:
\begin{equation}
f_i^{eq} = A_i + B_i (\bm{\xi} \cdot \mathbf{v}) + C_i |\mathbf{v} |^2 + D_i (\bm{\xi} \cdot \mathbf{v})^2 + \tilde{G}_i : (\bm{\xi} \otimes \bm{\xi}).
\end{equation}
Here coefficients $A_i, B_i,C_i,D_i,\tilde{G}_i$ are to be determined imposing conditions in Eqs.~\eqref{eqn:variables_hydro} and \eqref{eqn:constrain_second_moment}. In the continuum limit the Navier-Stokes equation is restored if $\eta=\tau/3$~\cite{denniston2001}.

\begin{figure}[t]
\centering
\includegraphics[width=.95\columnwidth]{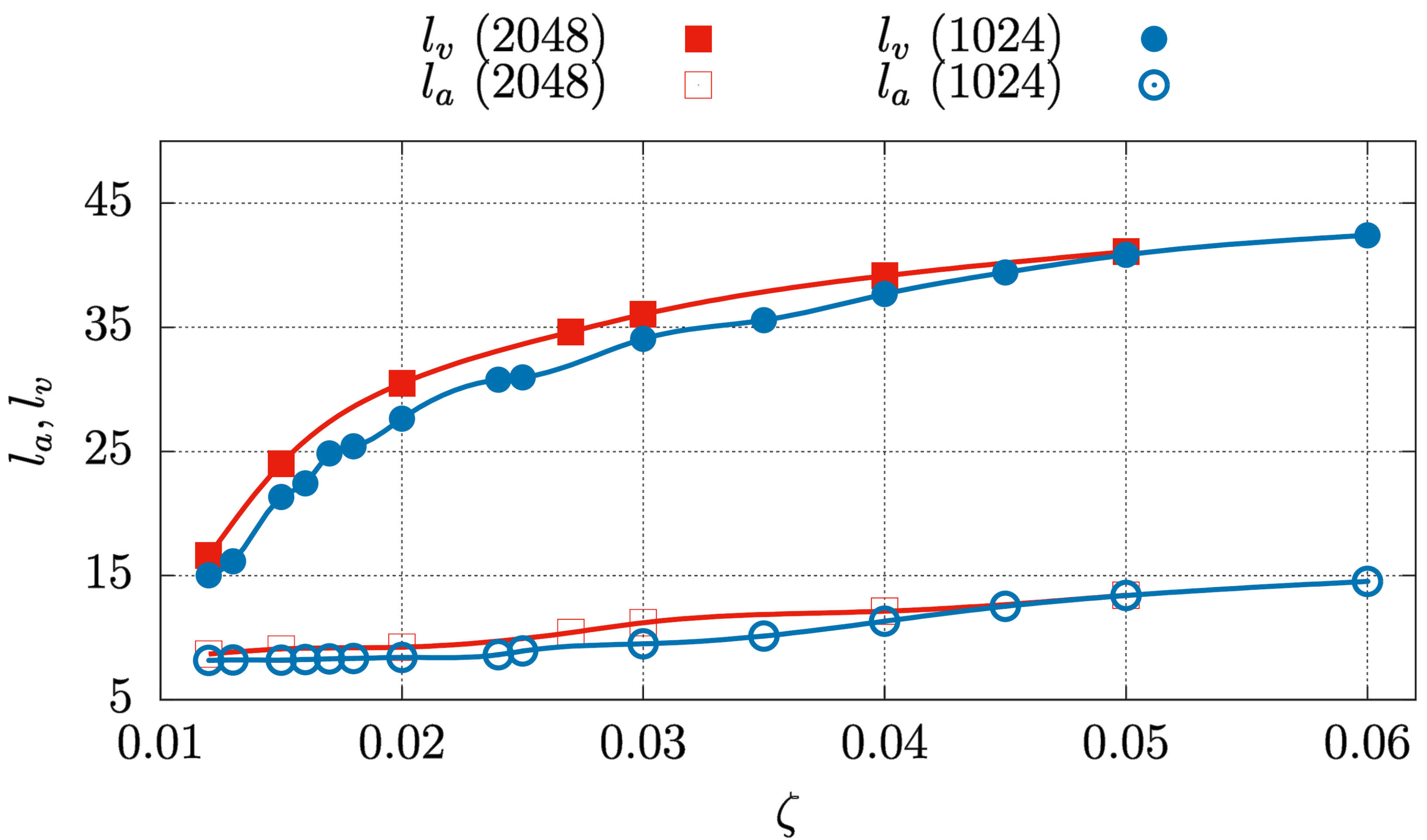}
\caption{Typical lengthscale of the velocity field $l_v$ and of the active injection $l_a$, as defined in the main text, as a function of $\zeta$ for systems of size $L=1024,2048$.}
\label{fig:fig1_SM}
\end{figure}

We performed simulations on bidimensional square lattices of size $L=1024,2048$ (Fig.~1,2,3,4 of the main text show results for a grid of size $1024$, while the dynamical response to the \emph{quench/unquench} of the activity parameter, shown in the inset of Fig.~4 of the main text, has been studied on a system of size $512$).
Periodic boundary conditions were imposed at the boundary. 
Table~\ref{table:table1} shows the simulation time and the Reynolds number $Re$, as defined in the main text, for the simulated cases at varying the activity parameter $\zeta$.
Fig.~\ref{fig:fig1_SM} shows a comparison between the active and velocity typical length-scales computed on grids of different size ($1024,2048$) to show that our results are not affected by finite-size effects.

\begin{table*}[]
\caption{Simulation details. The table shows for each value of the activity parameter $\zeta$ the simulation time in terms of LB iterations and the Reynolds number $Re$ measured at steady state as defined in the main text, both on a grid of size $L=1024$ and $2048$. n.s. stands for \emph{not simulated}.}
\begin{tabular}{|l|l|l|l|l|l|l|l|l|l|l|l|l|l|l|l|l|}
\hline
$\zeta$      & 0.012 & 0.013                 & 0.015 & 0.016                 & 0.017                 & \multicolumn{1}{l|}{0.018}                 & 0.020 & \multicolumn{1}{l|}{0.024}                 & 0.025                 & 0.027                 & 0.030 & 0.035                   & 0.040 & 0.045                 & 0.050 & 0.060                 \\ \hline
\multicolumn{17}{|c|}{1024}                                                                                                                                                                                                                                                                                                                              \\ \hline
time$/10^5$  & 42    & 42                    & 38    & 35                    & 30                    & \multicolumn{1}{l|}{30}                    & 30    & 30                                         & 30                    & \multirow{2}{*}{n.s.} & 30    & \multicolumn{1}{c|}{30} & 30    & 30                    & 30    & 30                    \\ \cline{1-10} \cline{12-17} 
$Re/10^{-2}$ & 4.8   & 5.6                   & 6.9   & 8.6                   & 12.0                  & 13.4                                       & 15.3  & 19.6                                       & 20.2                  &                       & 21.0  & 22.0                    & 26.3  & 33.3                  & 35.4  & 41.8                  \\ \hline
\multicolumn{17}{|c|}{2048}                                                                                                                                                                                                                                                                                                                              \\ \hline
time$/10^5$  & 40    & \multirow{2}{*}{n.s.} & 38    & \multirow{2}{*}{n.s.} & \multirow{2}{*}{n.s.} & \multicolumn{1}{l|}{\multirow{2}{*}{n.s.}} & 32    & \multicolumn{1}{l|}{\multirow{2}{*}{n.s.}} & \multirow{2}{*}{n.s.} & 28                    & 25    & \multirow{2}{*}{n.s.}   & 30    & \multirow{2}{*}{n.s.} & 25    & \multirow{2}{*}{n.s.} \\ \cline{1-2} \cline{4-4} \cline{8-8} \cline{11-12} \cline{14-14} \cline{16-16}
$Re/10^{-2}$ & 5.2   &                       & 7.4   &                       &                       & \multicolumn{1}{l|}{}                      & 18.2  & \multicolumn{1}{l|}{}                      &                       & 23.3                  & 25.2  &                         & 27.1  &                       & 35.1  &                       \\ \hline
\end{tabular}
\label{table:table1}
\end{table*}

For the analysis of the Poiseuille flow shown in Fig.~5  we made use of a grid of size $512 \times 256$. The system is confined between two horizontal flat walls at $y=0$ and $y=256$.
We imposed neutral wetting boundary conditions by requiring the following conditions to hold at the boundary:
\begin{equation}
\nabla_\perp \mu = 0, \qquad \nabla_\perp (\nabla^2 \phi)=0,
\end{equation}
where $\nabla_\perp$ denotes the partial derivatives taken along the normal to the walls. We imposed strong homeotropic  anchoring of the polarization field to the walls requiring:
\begin{equation}
P_\perp = 0. \qquad \nabla_\perp P_\parallel
\end{equation}
where $P_\perp$ and $P_\parallel$ respectively denote the normal and tangential component of the polariztion field with respect to the walls.

In our simulations the system is initialized in a mixed state, with the concentration field $\phi(\bm{r})= \phi_0/4	 +\delta\phi$, where $	\delta \phi$ is a random value uniformly distributed within the range $\left[ -\phi_{cr}/10, \phi_{cr}/10 \right]$, in order to obtain an asymmetric mixture with active and passive components respectively representing the 25\% and  75\% of the overall composition.
By varying the composition of the mixture, we checked that the hydrodynamic response is not altered, but for the value of $\zeta$ at which the transition (from the emulsified phase toward the mixed phase) takes place.
The polarization field is initially randomly oriented, while its modulus is randomly distributed between $0$ and $1$. The values of free energy parameters are $a=4\times 10^{-3}$, $\phi_0=2.0$, $\phi_{cr}=\frac{\phi_0}{2}$, $k_\phi=-6\times 10^{-3}$, $c=3 \times 10^{-2}$, $\alpha=10^{-3}$, $k_P=10^{-2}$, $\beta=10^{-2}$, while the constants appearing in the evolution equation have been chosen as follows: $M=0.1$, $\Gamma=1$, $\xi=1.1$. 
Notice that, by setting such value of $\xi$, the system is in the flow-aligning regime. Finally, in the case of Poiseuille flow, $f_x=5 \times 10^{-5}$.

\end{document}